\documentclass[aps,prl,twocolumn,superscriptaddress,longbibliography]{revtex4-1}

\usepackage{amsmath,amsfonts,amssymb}
\usepackage{braket}
\usepackage{graphicx}
\usepackage{tikz}
\usetikzlibrary{shapes,arrows,positioning,calc}
\usepackage{textcomp}

\def\centerarc[#1](#2)(#3:#4:#5){ \draw[#1] ($(#2)+({#5*cos(#3)},{#5*sin(#3)})$) arc (#3:#4:#5); } 

\newcommand{\mb}[1]{\mbox{\boldmath$#1$}}

\newcommand{\be}{\begin{equation}}
\newcommand{\ee}{\end{equation}}

\begin{document}

\title{Trapped electrons and ions as particle detectors}

\author{Daniel Carney}
\affiliation{Physics Division, Lawrence Berkeley National Laboratory, Berkeley, CA}
\author{Hartmut H\"{a}ffner}
\affiliation{Department of Physics, University of California, Berkeley, CA}
\author{David C. Moore}
\affiliation{Wright Laboratory, Yale University, New Haven, CT}
\author{Jacob M. Taylor}
\affiliation{Joint Quantum Institute, National Institute of Standards and Technology, Gaithersburg, MD}
\affiliation{Joint Center for Quantum Information and Computer Science, University of Maryland, College Park, MD}
\date{\today}

\begin{abstract}
Electrons and ions trapped with electromagnetic fields have long served as important high-precision metrological instruments, and more recently have also been proposed as a platform for quantum information processing. Here we point out that these systems can also be used as highly sensitive detectors of passing charged particles, due to the combination of their extreme charge-to-mass ratio and low-noise quantum readout and control. In particular, these systems can be used to detect energy depositions many orders of magnitude below typical ionization scales. As illustrations, we suggest some applications in particle physics. We outline a non-destructive time-of-flight measurement capable of sub-eV energy resolution for slowly moving, collimated particles. We also show that current devices can be used to provide competitive sensitivity to models where ambient dark matter particles carry small electric millicharges $\ll e$. Our calculations may also be useful in the characterization of noise in quantum computers coming from backgrounds of charged particles.
\end{abstract}

\maketitle

Detection of moving electrically charged particles is a ubiquitous problem in physics. Most charged particle detectors are based on ionization or scintillation of a detector medium, leading to detection thresholds on the order of eV. In the past few decades, semiconducting and superconducting devices based on lower energy excitations have been proposed with thresholds down to the meV range~\cite{hertel2019lhe,griffin2018polarphonons,hochberg2019nanowires,hochberg2016supercond}. However, for particles that interact via long-range forces like the Coulomb force, the predominant signal may occur in the center-of-mass or other collective mode of an object, which can have extremely low energies. As an example, several of us have recently proposed \cite{Carney:2019pza} and demonstrated \cite{Monteiro:2020wcb} the direct monitoring of impulses delivered to the motion of a macroscopic mass as a pathway toward ultra-low threshold detection in optomechanical systems \cite{Carney:2020xol}. 

Here we point out that analogous techniques can be applied to electrical interactions of charged particles with single trapped ions and electrons \cite{wineland1973monoelectron,paul1990electromagnetic,dehmelt1990experiments}. The achievable detection thresholds can be below the $\mu$eV scale. Since the electric force is long-ranged, even a single electron can offer substantial interaction cross-sections at sufficiently low detector thresholds, leading to remarkable detection reach. We focus primarily on electrons, since their extreme charge-to-mass ratio (at least $10^3$ higher than any ion) means that they set a fundamental sensitivity floor achievable with the known particles of nature. However,  the basic results carry over to ions by simply rescaling the electron mass $m_{e} \to m_{\rm ion}$. We also note that the same results would apply to any other long-range forces coupling to these systems, not just electromagnetism.

Electrons confined in Penning traps \cite{wineland1973monoelectron,dehmelt1990experiments} have a long history in metrology, notably in precision measurement of the electron magnetic moment \cite{hanneke2008new,hanneke2011cavity}. Electrons in both Penning and radio-frequency (rf) Paul traps have also been proposed as qubit systems which could enable very fast gate times in quantum computing \cite{ciaramicoli2003scalable,marzoli2009experimental,daniilidis2013quantum}, with single-electron trapping recently achieved experimentally in a room-temperature rf Paul trap \cite{matthiesen2021trapping}. This offers the exciting possibility of utilizing quantum information techniques to enhance the sensitivity of the device \cite{giovannetti2004quantum,riedel2013direct}.

As an example highlighting the extreme sensitivity of these devices, we consider the application of a single-electron sensor to searches for dark matter candidates with small electric charge $q \ll e$~\cite{mcdermott2011turning,PhysRevD.88.117701,Boddy:2018wzy,barkana2018strong,Emken:2019tni}. Such ``millicharged'' dark matter (mDM) can arise in models with weak mixing between the photon and hidden-sector forces~\cite{Holdom1986}, and has received significant recent interest to possibly explain \cite{barkana2018possible,berlin2018severely,barkana2018strong} an anomalous measurement in the 21 cm emission of cosmological hydrogen by the EDGES collaboration \cite{bowman2018absorption}. We show that a single electron in a typical trap configuration, utilizing only ground-state cooling and single-quantum (``single-phonon'') readout, would be capable of searching novel parameter space in viable scenarios in which $\sim1\%$ of dark matter consists particles with charges $\sim1\%$ of the electron charge.

\emph{Trapped electrons as impulse detectors.} Both Penning and Paul traps produce electron (or ion) motion which is well-described by harmonic oscillations. Penning traps use a static, axial magnetic field combined with a static electric quadrupole potential to trap the electrons. Paul traps instead do not use a magnetic field but rather a combination of a static electric quadrupole potential and rapidly varying drive field. Time-averaging over the drive produces dynamical stability. Electron motion in the Penning trap is characterized by the axial, cyclotron, and magnetron frequencies ($\omega_z,\omega_c,\omega_m$); Paul traps are characterized by three frequencies $\omega_z, \omega_x \approx \omega_y$ as well as the micromotion drive frequency $\Omega$. In either case, the frequencies are tunable; in Penning traps $\omega_m<\omega_z<\omega_c$.

We propose to exploit quantum control and readout of the electron motion as a sensitive impulse detector. Consider cooling the motion on an axis of our choice, with frequency $\omega$, to near the ground state $\ket{0}$. Suppose a rapid impulse is then delivered to the electron, imparting energy $\Delta E$. If $\Delta E \gtrsim \omega$, energy levels above the ground state will be populated. If some event, for example a particle collision, transfers momentum $\Delta p$ to the electron, this will impart energy $\Delta E = \Delta p^2/2 m_e$. Thus, the fundamental detection threshold of a single-electron detector is of order
\be
\label{sql}
\Delta p_{\rm SQL} = \sqrt{2 \hbar m_e \omega}.
\ee
This is precisely the ``standard quantum limit'' (SQL) for momentum detection using a harmonic oscillator \cite{mozyrsky2004quantum,clerk2004quantum,Carney:2020xol}. 
In terms of energy transfer, \eqref{sql} translates to a detection threshold simply given by the trap frequency. For electrons trapped in the 10 MHz-10 GHz range, this corresponds to threshold energies around neV-$\mu$eV, far below ionization levels. With ions, these frequencies can be down to the 10s of kHz (for example \cite{poulsen2012adiabatic}), which would produce similar momentum thresholds \eqref{sql} as the increased mass competes with the decreased frequency.

The essential duty cycle of a detector of this nature would proceed in three steps. First, we prepare the motion near the ground state $\ket{0}$. The system then evolves for some time $\Delta t$. After $\Delta t$ has passed, we estimate the change in momentum. A detector ``click'' occurs if this momentum change is greater than the threshold \eqref{sql} \footnote{We note that a detector of this type can be calibrated by a variety of methods. The most straightforward would be to use laser pulses of known shape \cite{Monteiro:2020wcb}; another would be to use a charged particle source (possibly with retarding potential to produce a slow beam)}.

We remark that the fundamental assumption we make in what follows is simply that we can measure changes in momentum at or near the SQL in a bandwidth similar to $\omega$. Thus, both the preparation and measurement elements can be relaxed to include higher temperature initial states and/or less effective measurement, so long as near SQL-limited momentum measurement is achieved. This has been demonstrated, for example, using both non-demolition \cite{hanneke2011cavity} and projective \cite{johnson2015sensing} measurement of the oscillator level. In addition to achieving measurement errors at SQL levels, the primary experimental requirement is that the heating rate (in phonons per second $\Gamma_Q$) should be small. Defining the quantum quality factor $Q = \omega/\Gamma_Q$, our duty cycle is bounded by $\Delta t \lesssim Q/\omega$, so that we will not obtain spurious clicks due to heating.

We substantiate these generic requirements by examining the cryogenic electron Penning trap in \cite{hanneke2011cavity}, in which the $150~{\rm GHz}$ cyclotron mode was monitored via quantum non-demolition measurements. This device satisfies all of the above requirements, and has $Q$ sufficiently high such that the effective ``duty cycle'' defined above can be on the order of days. Scaling this setup to lower-frequency modes (to lower the detection threshold \eqref{sql}) should be feasible. In a typical cryogenic ion trap with a $d \sim 50~{\rm \mu m}$ spacing between the ion and the electrodes, the phonon heating rates for modes around $1~{\rm MHz}$ are of order a few phonons per second \cite{Chiaverini2014a,Sedlacek2018-multi-mechanisms}, with a corresponding quantum $Q\gtrsim 10^5$. These rates should scale roughly as $1/d^4$~\cite{Brown2021} and linearly in $m_e/m_{\rm ion}$, so to achieve a rate of a few phonons per day would require building a trap with $d = \mathcal{O}({\rm cm})$. Typical dimensions of Penning traps approach this size and indeed the BASE collaboration finds heating rates of 2 quanta/day for the 18~MHz cyclotron motion of an anti-proton stored in a cryogenic Penning trap with $d=1.8$~mm~\cite{Borchert2019}.

\emph{Charged particle detection}. We now specialize to the case of a passing electrically charged particle $\chi$ of mass $m_{\chi}$ and charge $q_{\chi}$. The Coulomb force between this and the electron has magnitude $F = \lambda /r^2$. We will use ``natural'' units $\hbar = c = e = 1$, so $\lambda = \alpha q_{\chi}$ in terms of the fine structure constant $\alpha$.

\begin{figure}[t]

\includegraphics[scale=.8]{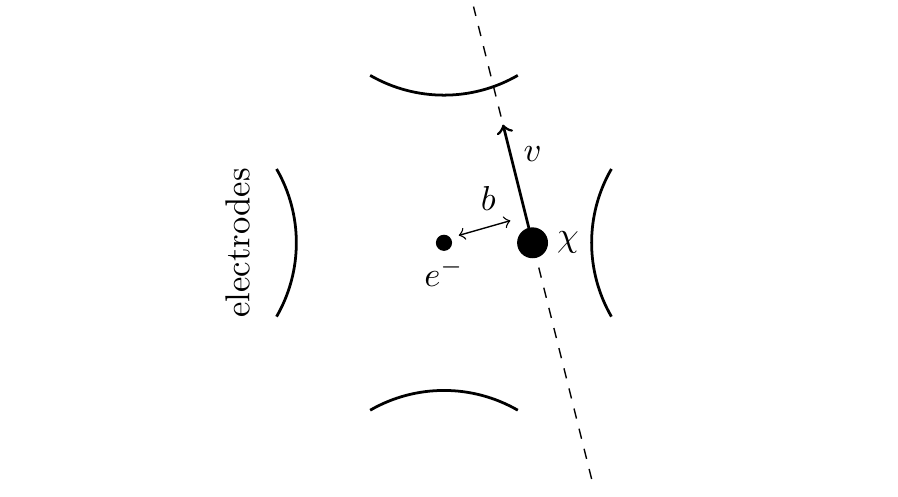}

\caption{Kinematics of a scattering event in a Paul trap. The charged particle $\chi$ impinges on the trapped electron with impact parameter $b$ and velocity $v$.}
\label{figure-kinematics}
\end{figure}

Let $b$ denote the impact parameter between $\chi$ and the electron and $v$ the relative initial velocity (see Figure \ref{figure-kinematics}). We can define a ``fly-by time'' $\tau = b/v$, during which the majority of the impulse is transferred. The simplest case, which we will assume here, is that $\tau \ll \omega_i^{-1}$ for all the trap frequencies (including, in the Paul trap, the micromotion drive frequency $\Omega$ \footnote{When $\tau$ is larger than $\Omega^{-1}$ but smaller than the inverse trapping frequencies $\omega_i^{-1}$ there may be important corrections to scattering due to the micromotion of the system, but we will assume Mathieu parameter $q_i \ll 1$, consistent with current experiments, and neglect this correction in what follows. We also note that axial detection will be insensitive to micromotion.}).  With the typical velocities and impact parameters of interest here, this should be valid for modes with frequency below around $100~{\rm GHz}$. In this limit, the electron is essentially a free particle during the collision, and we have simple Rutherford scattering. 

The events with the lowest momentum transfer will be at large impact parameter. This will produce small-angle scattering, for which the momentum transfer is
\be
\label{deltap-simple}
\Delta p = \frac{\lambda}{b v}.
\ee
We note that \eqref{deltap-simple} is true even for relativistic velocities $v \approx c$ \footnote{The 1st edition of Jackson's \emph{Electrodynamics} had a nice discussion of this (see his equation 13.1), but this material seems to have been removed in subsequent editions.}, although for sufficiently low velocity it breaks down because the incoming particle will not have enough kinetic energy.

We can combine \eqref{sql} and \eqref{deltap-simple} to estimate an effective cross-sectional area, $\sigma_{\rm eff}$, through which a passing particle will be detectable above threshold. Setting these equal and solving for the impact parameter gives
\begin{align}
\begin{split}
\label{aeff}
\sigma_{\rm eff} & = 4\pi b^2 = 4\pi x_0^2 \frac{\lambda^2}{v^2} \\
& \approx 40 ~{\rm nm}^2 \times \frac{q_{\chi}^2}{v^2} \times  \left( \frac{1~{\rm GHz}}{\omega} \right),
\end{split}
\end{align}
where $x_0^2 = 1/2 m_{\rm e} \omega$ is the size of the ground-state wavefunction of the electron. These cross-sections are sufficiently large to allow even a single trapped ion or electron to detect interesting fluxes of charged particles. 

These estimates assume that the momentum transfer due to scattering is entirely delivered to the mode of interest (say, the axial mode). For the Coulomb interaction, this will be the case only if the particle $\chi$ is incoming with velocity transverse to this axis. Thus if we know the direction of the incoming particles, this alignment is the optimum configuration. If we are instead trying to detect particles with unknown or random incoming directions, we will be sensitive only to a fraction of incoming particles
\be
\label{acceptance}
f_A(\Delta p) = \sqrt{1-\Delta p_{\rm th}^2/\Delta p^2}.
\ee
This natural (and tunable) anisotropy enables directional sensitivity of the detector and is thus a feature in applications like dark matter searches.

\begin{figure}[t]
\includegraphics[width=.9\columnwidth]{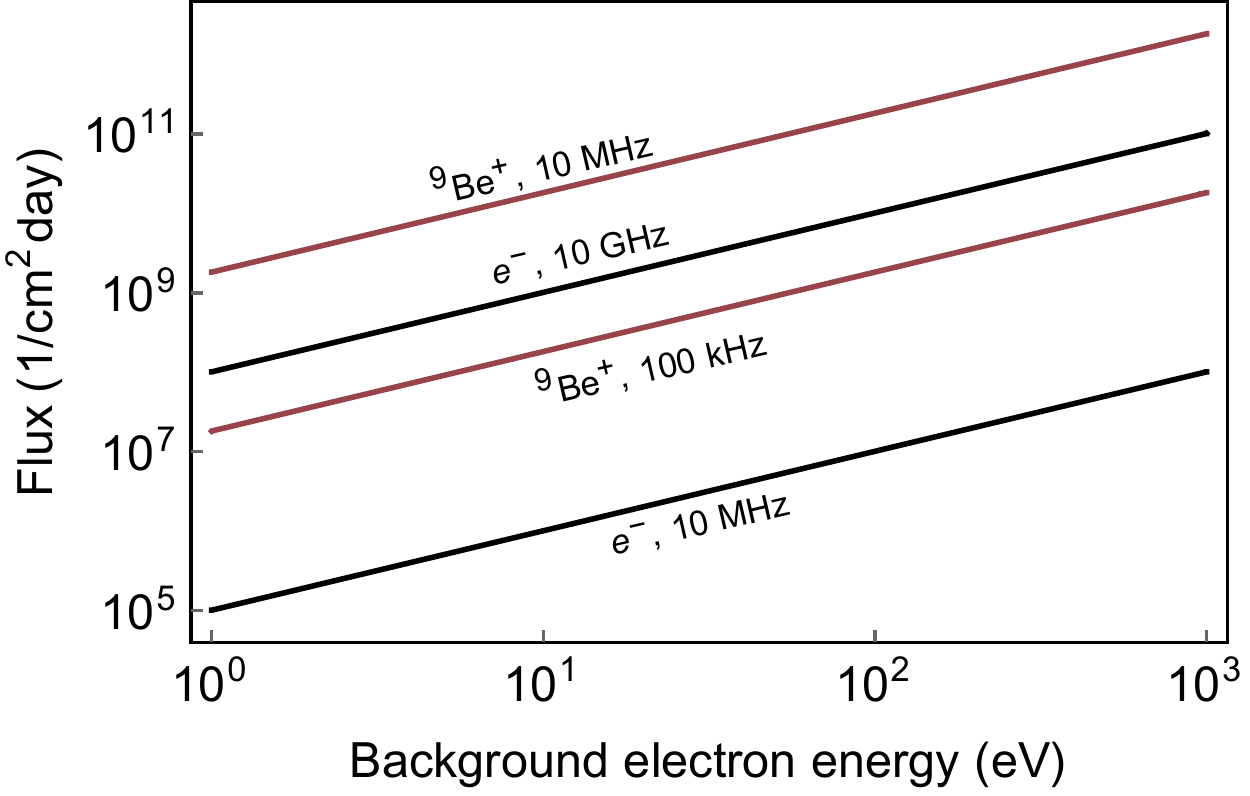}
\caption{Detectable levels of flux of ambient charged particles (here taken to be electrons), as a function of the ambient particle's kinetic energy. Each curve represents a trap with an electron or ion (here we use  beryllium \cite{biercuk2010ultrasensitive} to illustrate) with various frequencies. A background of electrons with the flux shown will produce impulses above threshold \eqref{sql} at a rate of 1 event/day.}
\label{plot-flux}
\end{figure}

As an illustration, Figure~\ref{plot-flux} shows the expected sensitivity for a trapped electron or ion to a given flux of background electrons. These could could arise for example from cosmogenic or radiogenic particles (e.g. the $\sim 10^3~{\rm cm}^{-2} {\rm day}^{-1}$ astrophysical muon flux) hitting the trap and causing showers of slow electrons. Alternatively, local sources of radioactivity may produce secondary low-energy particles at non-negligible fluxes, including secondary electrons or slow-moving recoiling daughter ions from decays on electrode surfaces.
While such radioactivity has been considered as a possible source of errors in superconducting qubits~\cite{Vepsalainen:2020trd,Mcewen:2021ood}, here we see that this may also eventually be a consideration for trapped electron or ion based systems. 

One possible application of such a detector could be to time-of-flight measurements of slowly moving charges in order to measure their kinetic energy \cite{jerkins2010using,steinbrink2013neutrino}. For example, one could use a retarding potential to slow electrons produced in nuclear reactions into a collimated beam. To resolve the energy of particles in the beam, one could use a coincidence measurement with a pair of trapped-charged devices spaced at distance $L$. The single-shot measurement time to resolve an SQL fluctuation is a single trap period $\Delta T = 2\pi/\omega$ \cite{clerk2004quantum}. A coincidence measurement would then be able to determine $v$ up to an error of order $\Delta v = L \Delta T/T^2 = v^2 \Delta T/L$. Resolving a $\chi$ particle's energy to precision $\Delta E = m_{\chi} v \Delta v$ would thus require a baseline of order $L = (2 E)^{3/2} \Delta T/m^{1/2} \Delta E$. For example, determining the energy of an $E \sim 1~{\rm eV}$ electron at a resolution of $\Delta E = 1~{\rm eV}$ could be achieved with a pair of $100~{\rm MHz}$ traps spaced at 10 mm. 

\emph{Sensitivity to millicharged dark matter (mDM)}. As a possible application of these ideas, we now consider detection of an ambient background of dark matter particles with small charges $q_{\chi} \ll 1$. Other DM-electron couplings---for example, coupling to electron number $L_e$ \cite{Arvanitaki:2016fyj}, or via a massive mediator \cite{emken2019direct}---would have similar detection reach, as long as the range of the force is longer than the cross section \eqref{aeff}.

Our detection target is thus an approximately homogeneous distribution of mDM particles, with positive and negative charges in arbitrary proportion. We assume these all have the same mass $m_{\chi}$. Astrophysical observations independent of the detailed nature of dark matter indicate that the average dark matter mass density should be around $0.3~{\rm GeV}/{\rm cm}^3$ \cite{bovy2012local}, so the number density of mDM particles is
\be
\label{nchi}
n_{\chi} =  \frac{0.3}{{\rm cm}^3} \times f_{\rm q} \times \left( \frac{1~{\rm GeV}}{m_{\chi}} \right).
\ee 
Here $f_q \lesssim 4 \times 10^{-3}$ \cite{barkana2018strong} is the fraction of DM which is charged. These charges will deliver random impulses on our detector. Let $f(\mb{v})$ denote their velocity distribution. The number of events, per unit time and  momentum transfer, is given by the Rutherford formula
\be
\label{dRdp}
\frac{dR}{d\Delta p} = n_{\chi} \frac{2\pi \lambda^2}{\Delta p^3} \eta(\Delta p).
\ee
Here $\eta(\Delta p) = \int_{v_{\rm min}}^{\infty} d\mb{v} f(\mb{v}) v^{-1}$ parametrizes effects from the velocity distribution, and $v_{\rm min} = \Delta p/m_e$ is the minimum incoming velocity kinematically necessary to produce a kick $\Delta p$ to the electron \cite{lin2019tasi}. Integrating \eqref{dRdp}, multiplied by \eqref{acceptance} for single-axis monitoring, from $\Delta p = \Delta p_{\rm th}$ upwards will produce the total number of events, per unit time, above the detection threshold $\Delta p_{\rm th}$. 

\begin{figure}[t]
\includegraphics[width=\columnwidth]{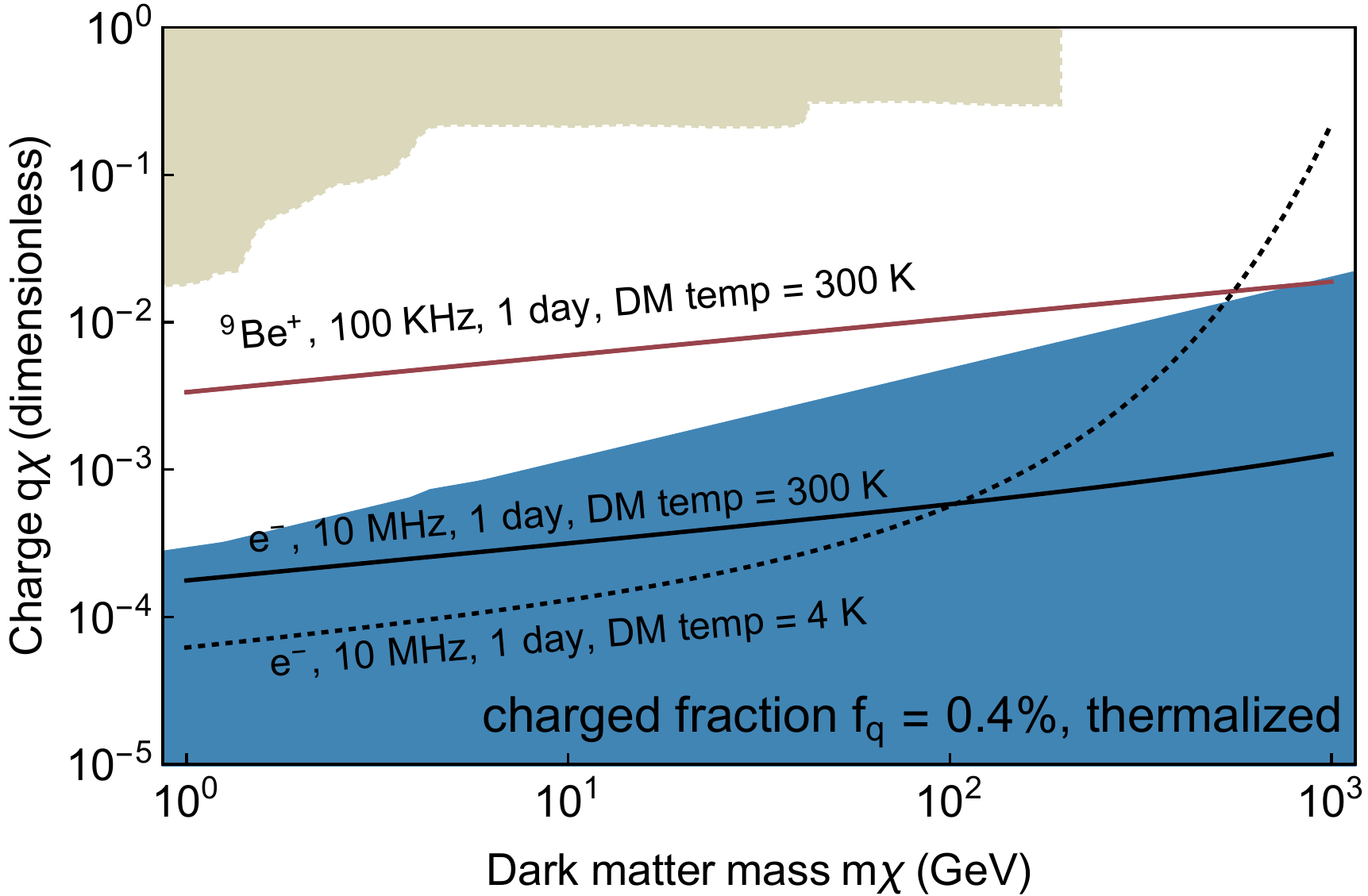} \\
\includegraphics[width=\columnwidth]{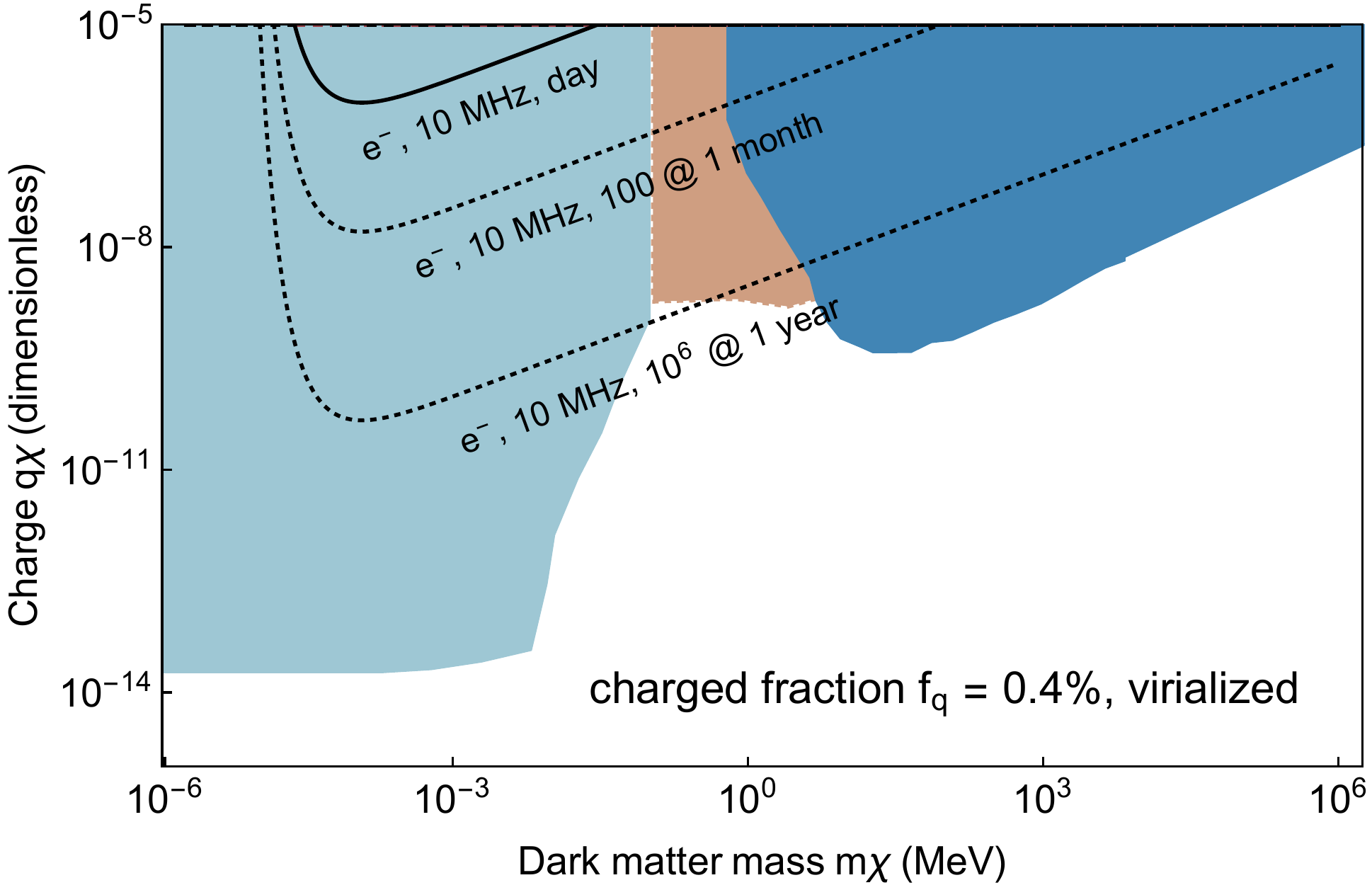}
\caption{Projected millicharged dark matter sensitivity with various trap frequencies and integration times. We consider detection along only a single trap axis \eqref{acceptance}. Top plot: high-charge DM thermalized at room temperature [300 K] (or at liquid He temperature [4~K]). Bottom: low-charge DM virialized to the galaxy. Currently excluded parameter space is shaded: bounds in light blue come from stellar astrophysics  \cite{boehm2013lower,chang2018supernova}, dark brown from cosmology (BBN and CMB $N_{\rm eff}$ \cite{boehm2013lower,vogel2014dark}), dark blue from direct detection experiments \cite{emken2019direct}, and light brown from a variety of collider experiments  \cite{prinz1998search,magill2019millicharged,acciarri2020improved,ball2020search}.}
\label{figure-constraints}
\end{figure}

For electrically charged DM, the details of the distribution present at a terrestrial experiment are poorly understood, since these charges interact with astrophysical magnetic fields, the atmosphere, etc. \cite{chuzhoy2009reopening,mcdermott2011turning,barkana2018strong}. In particular, at sufficiently large charge, mDM may lose substantial energy when transiting the atmosphere and earth en route to an experiment \cite{Emken:2019tni,Pospelov:2020ktu,Afek:2020lek}. Such energy loss explains the lack of direct detection limits above $q_{\chi} \gtrsim 10^{-4}$ \cite{emken2019direct}, because the subsequently slowed mDM will have kinetic energy below existing dark matter detection thresholds based on ionization \cite{Angle:2011th,Aprile:2016wwo,Agnes:2018oej} or excitation of electron-hole pairs in Si~\cite{Abramoff:2019dfb,PhysRevLett.123.181802damic,Amaral:2020ryn}. In contrast, the substantially lower thresholds possible with trapped electrons, as well as the the $v^{-2}$ enhancement in the cross section \eqref{aeff}, could enable detection in the currently unconstrained region of parameter space above the direct detection constraints around $10^{-5} \lesssim q_{\chi} \lesssim 10^{-1}$.

In Figure \ref{figure-constraints} we present a pair of example sensitivities in these two different charge regions. For charges $q \gtrsim 10^{-5}$ at relatively heavy mass $m_{\chi} \gtrsim 1~{\rm GeV}$, DM is expected to be captured and thermalized into the Earth \cite{Pospelov:2020ktu}. For smaller charges, in contrast, the DM should be virialized with the galaxy. In both plots we use the highly conservative number density \eqref{nchi} with $f_q = 0.4\%$; in particular, the number density for captured and thermalized mCPs may be orders of magnitude larger \cite{Pospelov:2020ktu}. Even with these conservative assumptions, we see that a single electron or ion monitored for a day has substantial detection reach in novel parameter space. 

Substantial improvements in cross-section could be reached with either multiple charges in a single trap, multiple traps, or a combination, as in quantum computing. For example, current ion traps can support crystals of $\gtrsim 10^2$ charges in a single trap spaced by their mutual Coulomb repulsion at $\gtrsim \mu$m distances~\cite{PhysRevLett.122.053603}. To illustrate the requirements to achieve sensitivity below the lower edge of cosmological constraints at $q_\chi \lesssim 10^{-10}$, we also show the sensitivity with $10^6$ electrons \footnote{While this may seem like a large number, note that this is an order of magnitude fewer qubits than required to factor a number relevant to RSA encryption with Shor's algorithm \cite{gidney2021factor}.}.

We note two key advantages of this type of detection. One is the inherent directional sensitivity. As described above, the trap has tunable anisotropy between the different axes, so directional sensitivity is possible for single events (by measuring multiple axes) or statistically with many events (by looking for temporal variations in the event rate in a single axis). Second, a number of traditional backgrounds are suppressed by the effective velocity discrimination in \eqref{aeff}.

Such background sources include the atmospheric muon flux described above, charged particles originating from plateout of Rn daughters (including from long-lived species such as $^{210}$Pb) on detector surfaces, and natural radioactivity such as $\beta$s from $^{40}$K in detector materials. These background sources have been studied in detail by rare-event searches for dark matter and $0\nu\beta\beta$ decay \cite{Akerib:2020com,Abgrall:2016cct,Leonard:2017okt}, and can be mitigated by material selection and handling to produce primary event rates sub-dominant to atmospheric muons. As described above in the context of qubit errors, production of low energy secondary charged particles remains an open question. Such secondaries have been identified in precision experiments sensitive to sources of low-energy $e^-$ such as KATRIN~\cite{Frankle:2011xy,Frankle:2020gxx,Aker:2021gma}. Nonetheless, given the small trap geometries, it is expected that it is feasible to reject a substantial fraction of such backgrounds by surrounding the traps with traditional charged particle detectors and vetoing coincidences with high-energy interactions.  Furthermore, the use of multiple electrons and searching for ``tracks'' would enable robust rejection of many backgrounds \cite{Carney:2019pza}. For these reasons, and the very low-event data from existing traps \cite{hanneke2011cavity,Borchert2019}, we assumed a background-free search is possible in producing Figure \ref{figure-constraints}, although more detailed study would be necessary in an experimental realization. 

In a very practical sense, our approach is to look for a sudden change in momentum due to an ``event'', when compared to stochastic and quantum noise backgrounds. The SQL is a reasonable starting point for the limits of this type of detection. We mention however that advanced techniques (e.g., using motional Fock states \cite{meekhof1996generation}, squeezing \cite{burd2019quantum}, or spin entanglement \cite{gilmore2017amplitude}) can push measurements beyond this limit. 

{\em Conclusions.} Single electrons have the highest charge-to-mass ratio known in nature. Like atoms, electrons are universal---each has exactly the same properties---and can be controlled and read out at the single-phonon level. Thus single-electron systems provide a fundamental platform for detection of tiny impulses, well below typical ionization energies. As a simple example, we showed that just one to a few trapped electrons can be used to perform novel searches for millicharged dark matter. Beyond the millicharged scenario, these systems can enable ultra low-threshold detection to any putative new long-range force coupling to the electron or ion. With trapped ions and electrons emerging as a platform for quantum information processing, these systems (with either one or many charges) also provide an opportunity to test quantum-enhanced protocols for sensing. 

\begin{acknowledgments}
\emph{Acknowledgements.} We thank Nikita Blinov, Gordan Krnjaic, Zhen Liu, Sam McDermott, Nadav Outmezguine, Ryan Plestid, Trey Porto, and Steve Rolston for discussions. DC is supported by the US Department of Energy under contract DE-AC02-05CH11231 and Quantum Information Science Enabled Discovery (QuantISED) for High Energy Physics grant KA2401032. HH acknowledges support from AFOSR through grant FA9550-20-1-0162, the
NSF QLCI program through grant number OMA-2016245. DCM is supported, in part, by NSF Grant PHY-1653232 and the Heising-Simons Foundation.
\end{acknowledgments}

\bibliography{references}

\end{document}